\documentclass[reprint,twocolumn,prl,superscriptaddress]{revtex4}
\newcommand{\pagenumbaa}{1}
\usepackage{graphicx}

\begin{document}

\title{Evidence for Helical Nuclear Spin Order in GaAs Quantum Wires}
\author{C.~P.~Scheller}
\affiliation{Department of Physics, University of Basel, Klingelbergstrasse 82, CH-4056, Basel, Switzerland}

\author{T.-M.~Liu}
\affiliation{Department of Physics, University of Basel, Klingelbergstrasse 82, CH-4056, Basel, Switzerland}

\author{G.~Barak}
\affiliation{Department of Physics, Harvard University, Cambridge, Massachusetts 02138, USA}

\author{A.~Yacoby}
\affiliation{Department of Physics, Harvard University, Cambridge, Massachusetts 02138, USA}

\author{L.~N.~Pfeiffer}
\affiliation{Department of Electrical Engineering, Princeton University, Princeton, New Jersey 08544, USA}

\author{K.~W.~West}
\affiliation{Department of Electrical Engineering, Princeton University, Princeton, New Jersey 08544, USA}

\author{D.~M.~Zumb\"uhl}
\email{dominik.zumbuhl@unibas.ch}\affiliation{Department of Physics, University of Basel, Klingelbergstrasse 82, CH-4056, Basel, Switzerland}

\begin{abstract}
\noindent
We present transport measurements of cleaved edge overgrowth GaAs quantum wires. The conductance of the first mode reaches $2\,e^2/h$ at high temperatures $T\gtrsim10\,$K, as expected. As $T$ is lowered, the conductance is gradually reduced to $1\,e^2/h$, becoming $T$-independent at $T\lesssim0.1\,$K, while the device cools far below $0.1\,$K. This behavior is seen in several wires, is independent of density, and not altered by moderate magnetic fields $B$. The conductance reduction by a factor of two suggests lifting of the electron spin degeneracy in absence of $B$. Our results are consistent with theoretical predictions for helical nuclear magnetism in the Luttinger liquid regime.
\end{abstract}

\maketitle
	
\setcounter{page}{\pagenumbaa}
\thispagestyle{plain}


Conductance quantization is a hallmark effect of ballistic one-dimensional (1D) non-interacting electrons \cite{Landauer1957,vanWees1988,Wharam1988,BeenakkerVanHouten}. One mode of conductance $e^2/h$ opens for each spin, giving conductance steps of $2\,e^2/h$ for spin degenerate electrons. In presence of electron-electron (e-e) interactions, strongly correlated electron behavior arises, described by Luttinger liquid (LL) theory \cite{Tomonaga1950,Luttinger1963,Haldane1981}. Salient LL signatures include ubiquitous power-law scaling \cite{Chang1996,Bockrath1999,Postma2001,Auslaender2001,Tserkovnyak2003}, separation of spin and charge modes, and charge fractionalization -- all recently observed \cite{Auslaender2002,Auslaender2005,Steinberg2008,Gilad2010} in cleaved edge overgrowth (CEO) GaAs quantum wires \cite{Pfeiffer1993}, thus establishing CEO wires as a leading realization of a LL. Interestingly, the conductance of a clean 1D channel is not affected by interactions, since it is given by the contact resistance in the Fermi liquid leads \cite{MaslovStone1995,Safi1995, Ponomarenko1995, Oreg1996, Picciotto2001}. In presence of disorder, however, the conductance is reduced with LL power-laws \cite{Tarucha1994, Maslov1995}. While short constrictions display universal quantization \cite{vanWees1988,Wharam1988}, the ballistic CEO wires exhibit steps reduced below $2\,e^2/h$ at temperatures $T\geq0.3\,$K \cite{Yacoby1996, Yacoby1997}, presenting an unresolved mystery \cite{Yacoby1996, Picciotto2000, Auslaender2001, Auslaender2002}.

In this Letter, we revisit the conductance quantization in CEO wires, investigating for the first time low temperatures down to $T\sim10\,$mK. We find that the conductance of the first wire mode drops to $1\,e^2/h$ at $T\sim100\,$mK and remains fixed at this value for lower $T$, while the electron temperature cools far below $100\,$mK. At high $T\gtrsim10\,$K, the conductance approaches the expected universal value $2\,e^2/h$ \cite{Yacoby1996}. This behaviour suggests a lifting of the electron spin degeneracy at low $T$, in absence of an external magnetic field $B$. The observed quantization values are quite robust, appearing in several devices, unaffected by moderate magnetic fields, and independent of the overall carrier density. A recent theory \cite{Braunecker2009PRB,Braunecker2010,Braunecker2009PRL} predicts a drop of the conductance by a factor of two in presence of a nuclear spin helix -- a novel quantum state of matter. Our data agree well with this model, while other available theories are inconsistent with the experiments, thus offering a resolution of the non-universal conductance quantization mystery.

\begin{figure}[ht]\vspace{0mm}
\begin{center}		
\includegraphics[width=8.3cm]{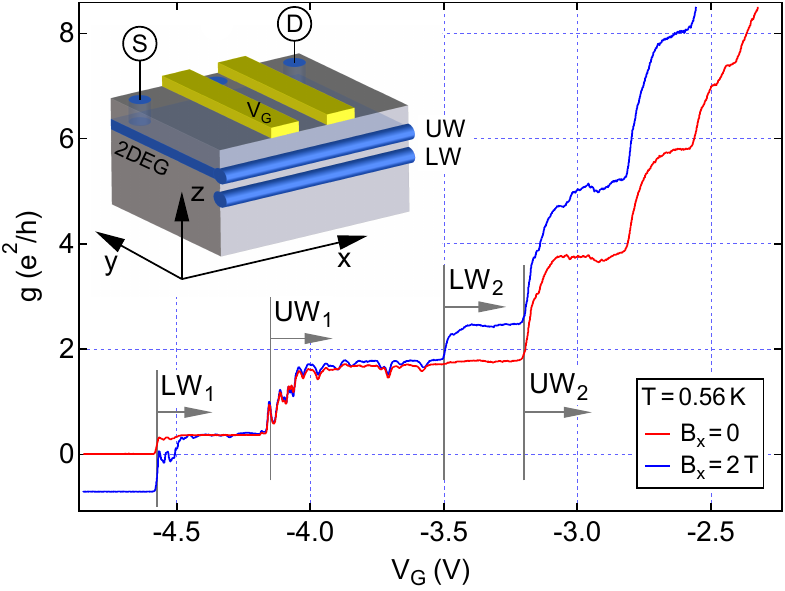}\vspace{-2mm}
\caption{{\bf Double wire mode structure} Differential conductance $g$ (red) versus gate voltage $V_G$ at $T=0.56\,$K and $B=0$. Arrows indicate $V_G$ above which modes start to contribute to $g$, as labeled. Blue data is at $B_X=2\,$T along the wire, offset in $g$ to align $LW_1$ plateaus. The inset shows a sample schematic with a coordinate system. }\label{wire_fig:1}
\end{center}\vspace{-6mm}		
\end{figure}

Ultra-clean GaAs CEO double wires (DWs) were measured (inset, Fig.\,\ref{wire_fig:1}), similar to Refs. \cite{Auslaender2002, Auslaender2005, Steinberg2008, Gilad2010}, offering mean free paths $\sim20\,\mathrm{\mu m}$ and subband spacings exceeding $10\,$meV. Details on sample fabrication are given in \cite{Pfeiffer1993,Yacoby1996,Yacoby1997, Auslaender2002}. A surface gate allows depletion of the 2D electron gas (2DEG) below, giving edge conduction in the DW only, forming what we will refer to as the ``wire''. Semi-infinite DWs with a few modes forming a 1D electron gas (1DEG) extend the wire on both sides, contacting the adjacent 2DEGs. Contacts to the 2DEGs are used to measure the two-terminal differential conductance $g$ of the wire.

The sample comprises an array of gates with $2\,\mathrm{\mu m}$ ungated spacing between $2\,\mathrm{\mu m}$ long wires, allowing individual and serial operation. In the ungated regions, the upper wire (UW) modes run directly adjacent to the 2DEG, resulting in a 2D-1D coupling length $\ell_{2D-1D}\sim6\,\mathrm{\mu m}$ \cite{Picciotto2000}. The 1DEG to few-mode wire transition occurs on a length scale of about $500\,$nm -- the distance of the UW and 2DEG to the surface gate -- clearly longer than the Fermi wavelength $\lambda_F\lesssim200\,$nm, and hence in the adiabatic regime. The lower wire (LW), on the other hand, has no adjacent 2DEG and is only weakly tunnel-coupled to the UW and 2DEG through a $6\, $nm thick AlGaAs barrier. UW to LW tunneling is very small in the gated segments. Thus, the $2\,\mathrm{\mu m}$ long DWs are considered as parallel resistors, with total conductance given by the sum of each conductance.

Figure \ref{wire_fig:1} allows identifying the wire modes as a function of gate voltage $V_G$: increasing $V_G$ starting from $g=0$ at the most negative voltages, $g$ is increasing in a step-like manner as the DW modes are populated one by one, as indicated. $LW_n$ ($UW_n$) denotes n-th mode in lower (upper) wire. Since the first step is small $\ll2\,e^2/h$, it is associated with the tunnel coupled $LW_1$.  The next, larger step corresponds to $UW_1$, followed by the $LW_2$ step, which becomes visible with a magnetic field $B_X=2\,$T along the wires (blue trace, shifted to align $LW_1$ plateaus). The tunneling process into the LW depends sensitively on parameters such as $B$, affecting the LW conductance. The next step has a large amplitude again and therefore corresponds to $UW_2$. Identifying higher modes is not easy due to a rapidly decreasing subband spacing.

\begin{figure}[t]\vspace{-0mm}
\begin{center}		
\includegraphics[width=8.3cm]{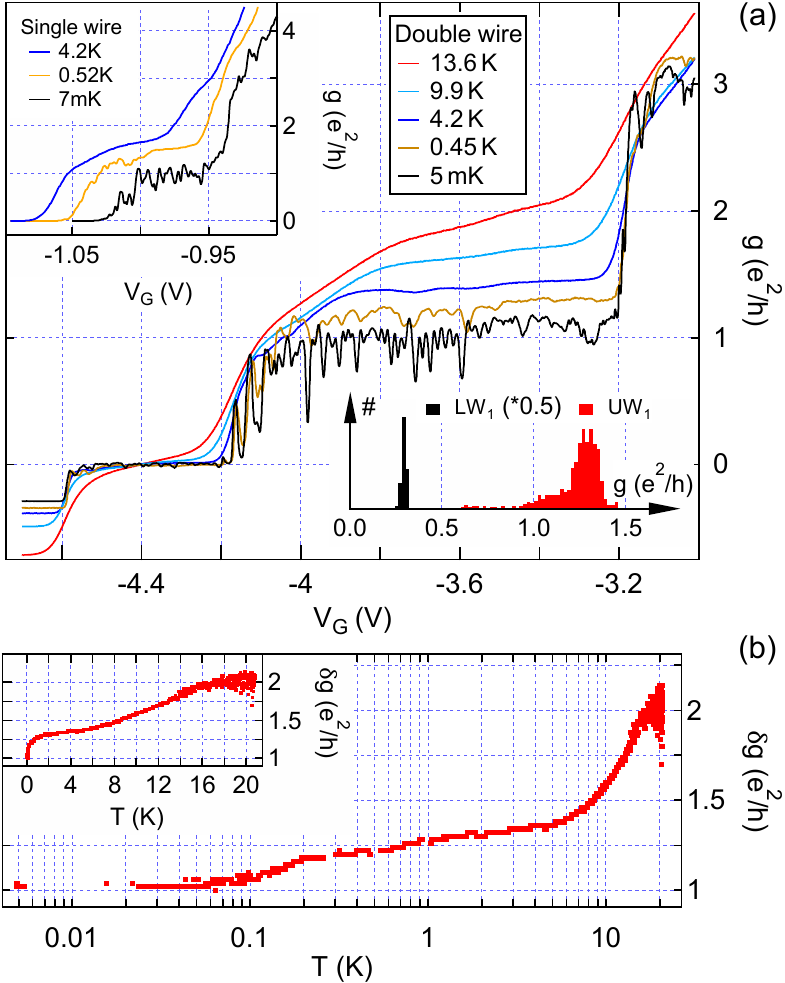}\vspace{-2mm}
\caption{{\bf Temperature effects} exhibiting conductance reduction by a factor of two (a) Gate voltage traces $g(V_G)$ at $T$ as labeled, shifted in $g$ to align $LW_1$ plateaus at $g=0$. Similar measurements for a single wire are given in the upper inset. Lower inset: histogram of $g(V_G)$ for $LW_1$ and $UW_1$ regions (base temperature). (b) Conductance step height $\delta g$ of $UW_1$ mode as a function of temperature on a logarithmic $T$-axis (linear axis in inset), extracted from histogram peak positions (see main text). Small but discrete steps in $g$ result from histogram binning.\vspace{-6mm}}
\label{wire_fig:2}
\end{center}		
\end{figure}

The temperature dependence is shown in Fig.\,\ref{wire_fig:2}. At high $T$, the $UW_1$ step height is approaching $2\,e^2/h$, as expected for a spin degenerate single mode wire. Thermally excited subband population and resulting inclined plateaus start to become visible at high $T$, as well as a feature reminiscent of $0.7$ structure \cite{Thomas1996} at the low end of the plateau. At low $T$, on the other hand, the $UW_1$ conductance plateau is reduced strongly to $\sim1\,e^2/h$, contrary to the $0.7$ feature, which rises to $2\,e^2/h$ at low $T$ \cite{Thomas1996}. In addition, the plateau develops pronounced, fully repeatable conductance oscillations. Very similar results are obtained for all four DWs, as well as for single wires, see upper inset in Fig.\,\ref{wire_fig:2}(a). Suitable single wires were not available (note the poor single wire plateaus). Hence the measurements were largely done on DWs.

The oscillation pattern on the $UW_1$ plateau -- complicating extraction of the step height -- is reproduced independent of the number of modes transmitted through an adjacent wire. This indicates ballistic addition of quantized mode steps, as expected for a mean free path far exceeding the wire length. The oscillations are well understood as quantum interference caused by the finite size of the wire \cite{paperFabry}, giving maximal transmission $\sim1$ at the conductance maxima. Indeed, the maxima of the oscillations neatly line up forming an upper ceiling on the $UW_1$ plateau, at intermediate $T$ even forming flat tops, see Fig.\,\ref{wire_fig:1}. The minima, on the other hand, are rather dispersed over a range of conductances. A histogram extending over the first two conductance plateaus clearly reflects this behavior, see lower inset Fig.\,\ref{wire_fig:2}(a). A long, asymmetric tail to low $g$ on the $UW_1$ plateau (red) is seen below the peak at higher $g$. Therefore, we extract the peak positions $g_{max\,UW_1}$ and $g_{max\,LW_1}$ from the histogram and obtain the $UW_1$ conductance step height $\delta g=g_{max\,UW_1}-g_{max\,LW_1}$.

The temperature dependence of $\delta g$ at $B=0$ is displayed in Fig.\,\ref{wire_fig:2}(b) from $20\,$K down to $5\,$mK. Starting from the highest $T$, where $\delta g$ reaches $2\,e^2/h$, lowering $T$ continuously and monotonously decreases $\delta g$ down to $\sim1\,e^2/h$. We note that breaking of spin degeneracy would result in a reduction of the conductance by a factor of two. At low $T\lesssim100\,$mK, $\delta g$ becomes temperature independent. However, the sample temperature cools far below $100\,$mK: first, thermal activation of fractional quantum Hall states can be used to extract an electron temperature $\leq27\,$mK, clearly smaller than $100\,$mK. Note that this $T$ is an upper bound only, since disorder can lead to deviation from exponential activation at low $T$. Occasional formation of a wire quantum dot \cite{Auslaender2001} leads to life-time broadened peaks not suitable for thermometry. Second, metallic Coulomb blockade thermometers \cite{Casparis2012} were measured under identical conditions, giving an electron temperature of $10.5\pm0.5\,$mK at refrigerator temperature $T=5\,$mK. Details on filtering and heat sinking will be given elsewhere \cite{Scheller2013}.

\begin{figure}[t]\vspace{1mm}
\begin{center}		
\includegraphics[width=8.3cm]{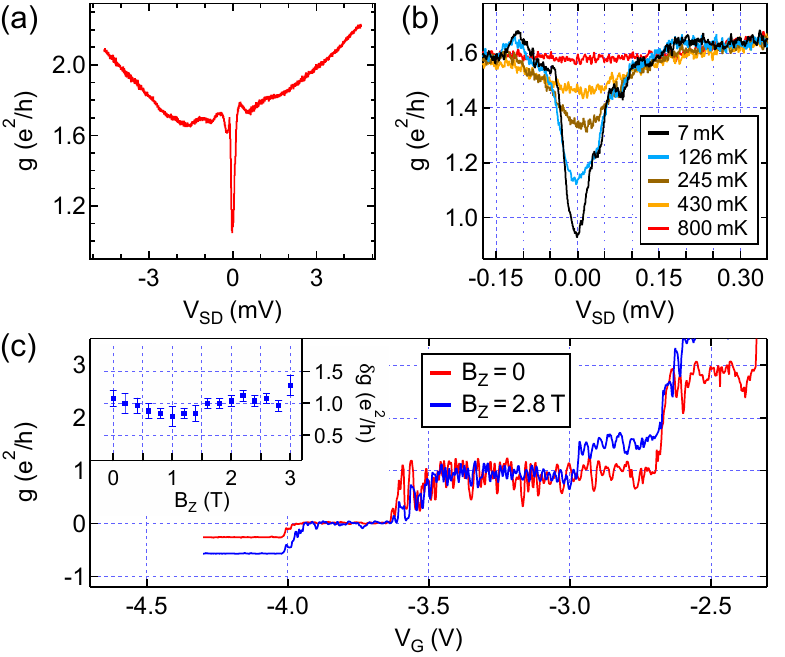}\vspace{-2mm}
\caption{{\bf Bias and B-field dependence} (a,b) $g$ as a function of dc bias $V_{SD}$, for fixed $V_G$ on the $UW_1$ plateau. (b) shows different temperatures as labeled. (c) $g$ at $B=0$ (red) and $B_Z=2.8\,$T applied perpendicular to the 2DEG (blue), shifted in $g$ to align $LW_1$ plateau. The inset shows the conductance step height $\delta g$ versus $B_Z$.\vspace{-6mm}}
\label{wire_fig:3}
\end{center}		
\end{figure}

Next, we investigate the dependence on source-drain bias $V_{SD}$. Fig.\,\ref{wire_fig:3}(a) and (b) shows the conductance $g_{UW}$ for $V_G$ fixed on the $UW_1$ plateau as a function of $V_{SD}$. $g_{LW}\sim0.3\,e^2/h$ depends only weakly on $V_{SD}$. At large $V_{SD}>1\,$mV, conductances around $2\,e^2/h$ are approached with rather weak $V_{SD}$ dependance, while at low $V_{SD}\sim0$, a sharp zero-bias dip of reduced $g_{UW}\lesssim1\,e^2/h$ develops. Thus, the $|V_{SD}|$ behavior and the $T$-dependence appear qualitatively very similar. The zero-bias anomaly corresponds to the steep drop of $g(T\rightarrow0)$ clearly visible in the inset of Fig.\,\ref{wire_fig:2}(b) when plotting $\delta g$ on a linear $T$ axis. Given the very strong $g(V_{SD})$ dependence, great care was taken to keep $V_{SD}$ very small throughout all linear response measurements.

\begin{figure}[t]\vspace{-1mm}
\begin{center}		
\includegraphics[width=8.3cm]{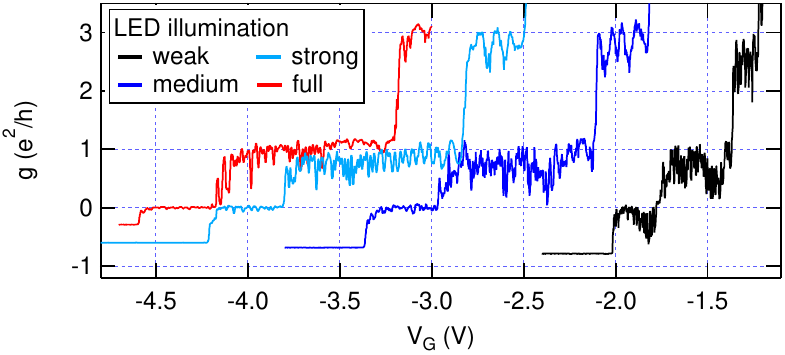}\vspace{-2mm}
\caption{{\bf Density dependence} $g(V_G)$ for a DW, recorded after LED illumination as labeled. Traces were shifted in $g$ only to align $LW_1$ plateau at $g=0$. $\delta g$ appears independent of flash strength and hence carrier density.\vspace{-6mm}}\label{wire_fig:4}
\end{center}		
\end{figure}

We now turn to the influence of a magnetic field. Fig.\,\ref{wire_fig:3}(c) compares $g$ at $B=0$ and $B_Z=2.8\,$T perpendicular to the 2DEG, at base $T$. While $g_{LW}$ is changed, the step height $\delta g$ is hardly affected at all: $\delta g(B_Z)$ remains close to $1\,e^2/h$ within the error bars (see inset Fig.\,\ref{wire_fig:3}(c)), despite Landau levels and edge states induced by $B_Z$ in the 2DEG, reaching filling factor $\nu=3$ at $B_Z=3\,$T. Further, the transitions from $LW_1$ to $UW_1$ at the larger $B$ are comparable to $B=0$ data (see e.g. Fig.\,\ref{wire_fig:4}) and do not provide evidence for an additional plateau. Note that at $3\,$T, the Zeeman splitting is much larger than temperature, and the Landau level spin splitting is already resolved for much lower $B_Z\sim0.3\,$T. Finally, $\delta g$ shows very little  dependence on $B_X$ (Fig.\,\ref{wire_fig:1}). Overall, we did not find evidence for qualitative changes of the $UW_1$ conductance step in moderate $B$-fields.

We emphasize that the experiments \cite{Yacoby1996,Picciotto2000}, which studied single wires at $T\geq300\,$mK, are consistent with the results presented in this Letter. New here is the full $g$ reduction to $g\sim1\,e^2/h$, $T$-independent for $T\lesssim100\,$mK, combined with the sharp zero-bias dip, $B$-field independence, and pronounced low $T$ conductance oscillations. In light of our new and complementary data, we now proceed to analyze different theories attempting to explain our findings, including re-examining models already discussed in Refs.\,\cite{Yacoby1996, Picciotto2000}. First, non-interacting theories must be rejected: reduced conductance quantization within the Landauer formula results from non-ideal transmission $t<1$ \cite{BeenakkerVanHouten}, in contradiction to the observation of ballistic transport in our wires, in addition to the objections already raised in Ref.\,\cite{Yacoby1996}. Our measurements with two wires in series show that $t$ is at most a few percent below $t=1$, in any case ruling out a $g$-reduction by a factor of two.

Second, we examine e-e interactions in the wire. A weakly disordered LL connected to Fermi liquid (FL) leads \cite{Maslov1995} gives conductances decreasing below $2\,e^2/h$ with a power-law in $V_{SD}$ and $T$. A finite conductance $\propto L^{-1}$ is obtained at $T=0$ due to thermal freeze-out: when the thermal length $L_T$ exceeds the wire length $L$ at low enough $T$, $g$ becomes $T$ independent. However, here, $\delta g(T)$ remains clearly $T$ dependent well below the freeze-out temperature $\sim 0.6\,$K (see Fig.\,\ref{wire_fig:2}) and further cannot reasonably be fit with a single power law over the entire $T$-range. Therefore, LL theory for the $2\,\mathrm{\mu m}$ wire alone is an unlikely explanation.

Next, we consider e-e interactions also outside the wire. The 1DEGs may also experience non-FL correlations, albeit weaker than the wire since the 1DEGs are not single mode. The 2D-1D coupling scale $\ell_{2D-1D}\sim\,6\,\mathrm{\mu m}$ sets an effective LL system length $L_{1DEG}=2\cdot \ell_{2D-1D}+L$ comprised of segments $\ell_{2D-1D}$ on each side of the $L=2\,\mathrm{\mu m}$ wire. As $T$ is reduced, $L_T$ first grows larger than $L$ before eventually surpassing $L_{1DEG}$, where $g(T\rightarrow0)$ saturates at $g_{sat}\propto 1/L_{1DEG}$. Hence, two temperature ranges with distinct power laws emerge, before $g$ saturation at low $T$.

$\delta g(T)$ is consistent with such a model, giving decent agreement with two separate power-law fits. Further, a reasonable saturation temperature results: $L_T>L_{1DEG}$ occurs on a temperature scale of $\sim0.1$K, where indeed the $\delta g$ data is seen to loose $T$-dependence. The value $g_{sat}\sim1\,e^2/h$ could then simply be a coincidence, but would depend on the details of the 2D-1D coupling. This coupling must involve scattering at an impurity or defect due to the large momentum mismatch between 1DEG and 2DEG electrons \cite{Yacoby1996}, and hence, within this model, $g_{sat}$ will depend on parameters \cite{Yaroslav} such as disorder, chemical potential (density), and $B$-fields.

Figure\,\ref{wire_fig:4} displays $g(V_G)$ for a sequence of LED illumination \cite{Auslaender2002} steps, ionizing more and more donors and thereby globally increasing the carrier density and mobility after each flash. The depletion voltage is proportional to density, and is seen to become more negative with increasing LED exposure, enhancing the density by over a factor of two, see Fig.\,\ref{wire_fig:4}. Similarly, the 2D density and mobility increase by roughly a factor of two (before illumination, the density is $1\cdot10^{11}\,\mathrm{cm^{-2}}$ and the mobility $\sim3\cdot10^6\,\mathrm{cm^2/(Vs)}$). Despite the large density change, the $UW_1$ step height (ceiling of $g$ oscillations) is seen to remain very close to $1\,e^2/h$. This is seen also in the other wires. In absence of any significant density, disorder, wire, and B-field dependence (see Fig.\,\ref{wire_fig:1} and \ref{wire_fig:3}) of $g_{sat}$, this scenario has to be abandoned.

A further model put forth in \cite{Yacoby1996} and refined in \cite{Picciotto2000} proposed a competition between $\ell_{2D-1D}$ and residual backscattering in the wires on a length $\ell_{BS}\gg L$ for the reduced $g$ plateaus. This model is expected to exhibit a similar sensitivity to the 2D-1D coupling details as above, and can again be ruled out based on the observations in Fig.\,\ref{wire_fig:4}, augmenting objections already raised in \cite{Yacoby1996, Picciotto2000}. In addition, both $\ell_{2D-1D}$ and $\ell_{BS}$ have (weak) LL power-law $T$-dependence \footnote{$\ell_{2D-1D}\propto T^{-p_1}$ and $\ell_{BS}\propto T^{p_2}$, with $0<p_{1,2}<1$.}, leading to $g\rightarrow 0$ for $T\rightarrow 0$, in contradiction to the finite $g_{sat}$ observed. Another scenario is an incoherent LL due to Wigner crystal formation \cite{Matveev2004PRL,Matveev2004PRB}. In this model, $g$ increases from $1\,e^2/h$ to $2\,e^2/h$ upon decreasing temperature, opposite to observations here. Further, very low densities $(a_B n)^{-1}\gg1$ are required ($a_B$ is the GaAs Bohr radius), which is not the case for the wires used here. Finally, spin orbit coupling has to be ruled out as well, since the $g$-reduction is seen at $B=0$ and shows little $B$-dependence.

A recent theory by Braunecker, Simon and Loss \cite{Braunecker2009PRB,Braunecker2010,Braunecker2009PRL} predicts helical nuclear spin order in a LL, causing a reduction of $g$ by a factor of two, from $2\,e^2/h$ to $1\,e^2/h$ for a clean wire, as seen in the experiment here. Below a crossover temperature $T^*$, an effective RKKY interaction, strongly enhanced by e-e interactions, forces the nuclear spin system via hyperfine interaction into helical order, constituting a novel state of matter. The resulting large Overhauser field acts back on the electronic system where a large gap opens -- pinned at the Fermi energy -- for half of the low energy modes, forming a helical LL and causing the $g$ reduction at $B=0$, applicable similarly for single and double wires \cite{Meng2013two}. The wire then only transmits spin-down right and spin-up left movers, therefore acting as a perfect spin filter. Note that the nuclear spin helix is a thermodynamic ground state protected by a gap, rather than a dynamic nuclear spin polarization.

The predicted $T^*$ depends very strongly on the charge LL parameter $K_C$ and can exceed $1\,$K for small $K_C$ (strongly interacting) \cite{Braunecker2009PRB}. Full nuclear order is obtained only at $T\ll T^*$ and zero polarization only at $T\gg T^*$. Estimating $K_C$ is far from trivial both experimentally and theoretically \cite{Auslaender2005}: $K_C=0.4$ gives $T^*\sim0.2\,$K and $K_C=0.3$ already $T^*\sim0.6\,$K, consistent with the experiment. Further, large $T^*$ result in a rather broad, washed-out transition, as observed in the experimental $\delta g(T)$. $K_C$ is expected to depend (weakly) on density $n$, therefore $T^*$ will change over a conductance plateau. However, given a very broad transition, this may affect $g$ only weakly, and give rather flat conductance plateaus, as seen in the experiment. Further, the theory derives $g$ far below and above, rather than throughout, the nuclear transition, allowing only a qualitative comparison. Finally, a Zeeman splitting much smaller than the induced gap should affect neither the nuclear order nor the conductance, as seen in the experiment.

In summary, we have investigated conductance quantization in single mode LL wires, finding a very broad transition from $2\,e^2/h$ at high $T$ to $1\,e^2/h$ at low $T\lesssim\,100\,$mK, where $g$ becomes $T$ independent and shows pronounced $g$ oscillations.  This behavior is consistently seen in several double and single wires, is independent of illumination, and is insensitive to moderate $B$ fields. These data are in good agreement with a recent nuclear spin helix model \cite{Braunecker2009PRB,Braunecker2010,Braunecker2009PRL}. All other theories considered here and previously \cite{Yacoby1996,Picciotto2000} are inconsistent with the experiment. Thus, the present results offer first evidence for a nuclear spin helix in a GaAs quantum wire.

Magnon spectroscopy might give direct evidence for nuclear spin involvement and helical spins in particular. Resistively detected NMR was already attempted here: while detecting clear 2DEG signals, no identifiable NMR response was found for the wires. This is consistent with the notion that flipping spins in a gapped helix would be energetically unfavorable. Further, spectroscopic methods \cite{Auslaender2002} might be used to shed more light on the electronic structure. In the nuclear spin helix state, the electron system is in the helical LL regime, equivalent to a spin-selective Peierls transition in a Rashba spin-orbit coupling wire \cite{Braunecker2010}. Given proximity to an s-wave superconductor, a topological phase sustaining Majorana fermions could be created.

\begin{acknowledgments}
We would like to thank O.~Auslaender, B.~Braunecker, D.~Loss, D.~L.~Maslov, T.~Meng, M.~Meschke, J.~Pekola, P.~Simon and Y.~Tserkovnyak for valuable inputs and stimulating discussions. This work was supported by the Swiss Nanoscience Institute (SNI), NCCR QSIT, Swiss NSF, ERC starting grant, and EU-FP7 SOLID and MICROKELVIN. AY acknowledges support from the NSF DMR-1206016.
\end{acknowledgments}



\end{document}